\documentclass{elsart}
\usepackage[dvips]{graphicx}

\begin{document}
\begin{frontmatter}

\title{Imaging Gaseous Detector based on Micro Processing Technology}
\author{Toru~Tanimori, Yuji~Nishi, Atsuhiko~Ochi, Yasuro~Nishi}
\address{Department of Physics, Tokyo Institute of Technology, Tokyo 152, Japan}

\begin{abstract}

The development of  gaseous detectors has been
exciting again since the appearance of a MicroStrip Gas Chamber(MSGC)
in 1988, which is made using a micro-electronics technology.
These days lots of variations of the advanced gaseous detectors 
are being intensively studied in the world. 

We have developed the two-dimensional MSGC having a 10 cm 
square detection area and the ultra fast readout system
for a real time X-ray imaging. 
The MSGC was made using Multi-Chip Module (MCM) technology, and
has a very thin substrate of 17 $\mu$m, lots of thin anodes and
back strips both with 200 $\mu$m pitches.
This enables us to get  fast digital X-ray images
with recording both the timing and an energy of each X-ray photon.  
In addition, an intermediate gas multiplier 
has been realized using a capillary plate having a
conductive surface of a capillary.
The MSGC combined with the conductive capillary plate 
can be  steadily operated with a  high gain  under intense irradiation. 
Here we also report new approaches of X-ray crystal structure analyses
using timing information obtained by the MSGC.
\end{abstract}


\end{frontmatter}

\maketitle

\section{Introduction}
The MicroStrip Gas Chamber (MSGC) was 
proposed in 1988 by Oed [4].
This new detector has been expected to provide 
stable  operation under intense irradiation
and  radiation hardness, which are 
required for an X-ray photo-counting detector operated in
high intensity radiation environments.
MSGCs also have a good position resolution of $\sim$ a few tens $\mu$m.
Therefore a two-dimensional MSGC would enable  
to realize an ideal X-ray imaging detector also  
having a photo-counting ability.
  
A MSGC is usually produced using micro-electronics technology: 
sequences of alternating thin anodes and cathodes are formed  with a 
few hundred micron pitch  on an insulating substrate. 
The closeness of  the electrodes provides the above features of a MSGC.
While most MSGCs have been  realized on glass or quartz substrates so far, 
we have been developing  another type of  MSGC having a $\sim$20$\mu$m thin 
polyimide substrate since 1991(Nagae et al.[2], Tanimori et al.[8]). 
Our MSGC is made using Multi-Chip Module (MCM) technology, 
which allows a  high density assembly of bare silicon LSI chips on a silicon 
or a ceramic board.  

A very thin substrate of the MSGC 
enables us to control the flow of positive ions from anodes to cathodes
by optimizing the potential on the back plane. 
Also due to the thin substrate,
a fast signal is induced on a backplane, which enables two-dimensional 
readout from one MSGC (Nagae et al.[2], Tanimori et al.[7]).
In 1993, the 5cm square 2D-MSGC 
with  200 $\mu$m anode and backstrip pitches  were made, and
clear two-dimensional X-ray images were successfully  obtained.
The performances obtained from this MSGC were 
described  in detail in Tanimori et al.~[8],
in which position resolution, stability, durability, and operation in
high counting rate were investigated.

Based on that study,
the new 2D-MSGC having a large area of a 10cm square has been developed 
since 1997.
In addition, we have developed a new type of the readout system
in which the data are synchronously managed in digital electronics
to handle the huge quantity of data from the MSGC for real-time
image processing.
Using this new system,  we successfully got real-time movies from the MSGC, 
and  examine the new X-ray crystal analysis methods
using the photo-counting ability of the MSGC (Tanimori et al. [9]).

Here, after summarizing both the MSGC imaging device and the readout system,
we mainly report the new approach to overcome the destruction 
of electrodes due to discharges, which is the most crucial problem for MSGCs,
and the further development of the application of the MSGC imager 
to the X-ray crystal analysis.

\section{Structure of the MSGC}

\begin{figure}
\begin{center}
\includegraphics[width=120mm]{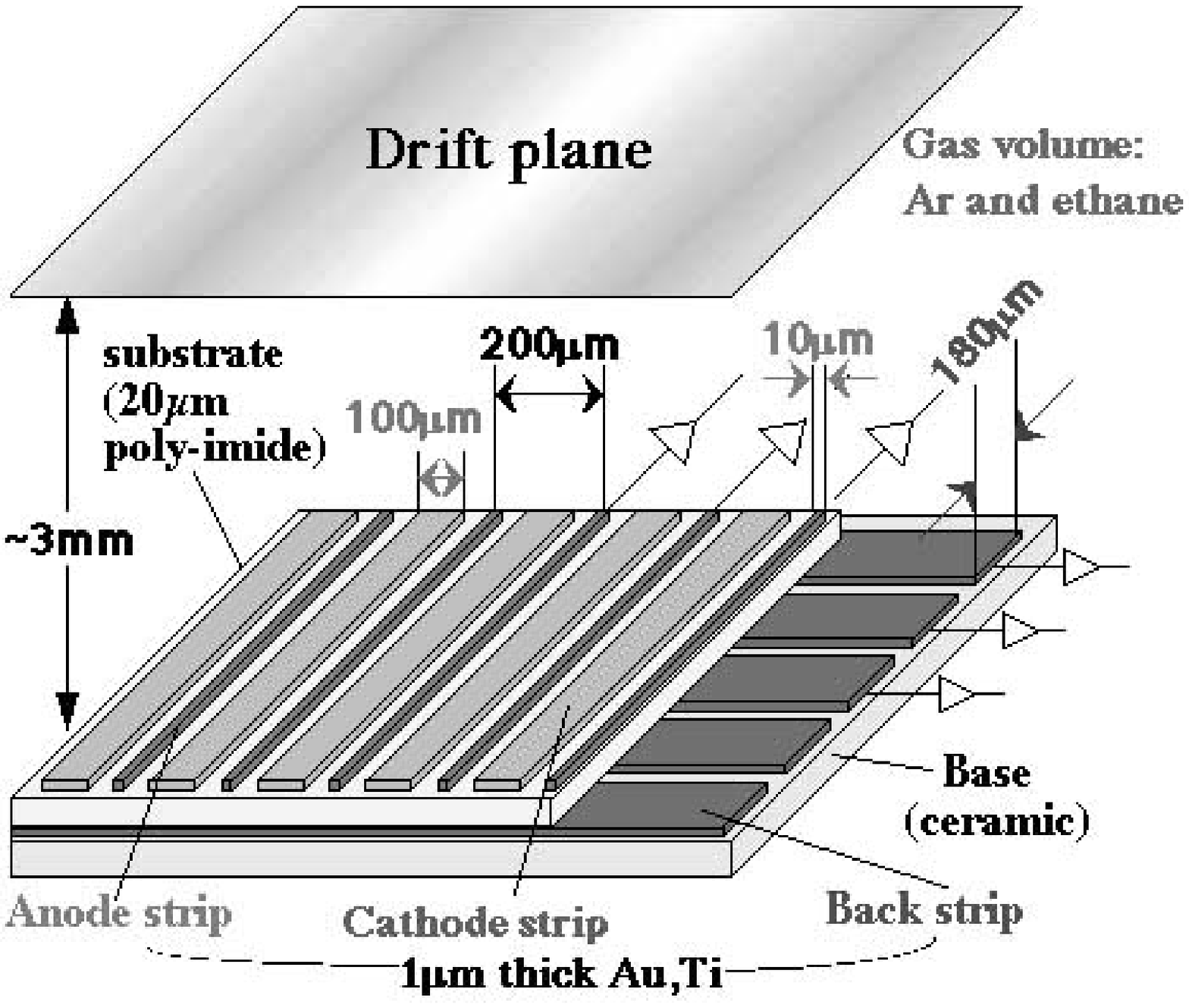}
\end{center}
\caption{}
\label{fig:f1}
\end{figure}

Figure \ref{fig:f1} 
shows a schematic structure of our two-dimensional 
MSGC, which is formed  on the 20 $\mu$m thin 
polyimide substrate.
On the polyimide layer, 10 $\mu$m wide anodes  
and 100 $\mu$m wide cathodes 
are formed alternately by  photo-lithography technology.
Between the ceramic base and the polyimide substrate,
there are back strips with  a 200 $\mu$m pitch 
in orthogonal to the anode,
which provide information in the second dimension. 
All electrodes are made of gold with a thickness of 1 $\mu$m
(recently chromium are used as mentioned in section 5). 
In order to reduce the effect of parallax broadening
of the position distributions,
a drift  plane is placed 3 mm above 
the substrate.
Every 32 cathode strips  are aggregated to one group at one end in the 
a 10cm square MSGC.
There are 16 cathode groups, to which high voltages are independently supplied.
The signal from groups of cathodes can 
be used  as an  energy measurement of an X-ray.
The edge of the cathode is coated by polyimide with a width of $\sim$7 $\mu$m 
for suppressing discharges between anodes and cathodes(Tanimori et al.[8]) .  
The back strips are connected to the preamplifiers. 
A gas mixture of Argon (80\%) and  C$_2$H$_6$ (20\%) was used in atmospheric pressure.
The absorption efficiency  of this condition 
(3mm gap and the use of the above gas) is $\sim$8\% for 8.9 keV X-rays.
The resistivity of the substrate  is considered to be a key 
factor to keep stable operation of a MSGC under high counting rate.
A very thin organic-titanium is coated on the surface,
by which a surface resistivity of $\sim 10^{15}$ $\Omega$/square 
was obtained.
We found an optimum operating point by adjusting 
both the thickness of the substrate
and the potential of both anodes and cathodes.
The details of the features about gas amplification and its stability
of the MSGC are reported in Tanimori et al.~[8].

\section{Read-out Electronics System}
   
\begin{figure}
\begin{center}
\includegraphics[width=100mm]{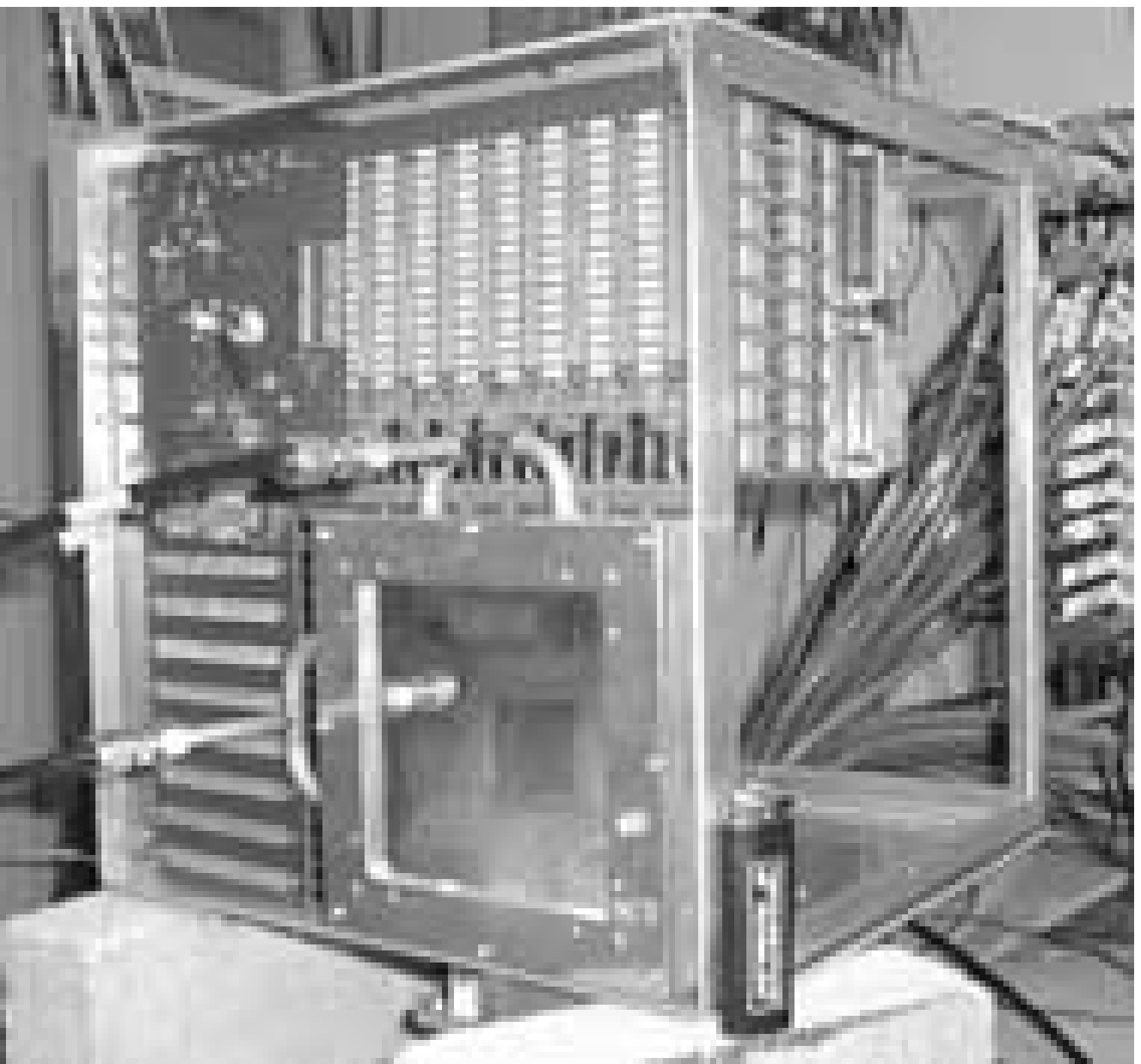}
\end{center}
\caption{}
\label{fig:f2}
\end{figure}

The  new 10 cm square MSGC was directly mounted on 
the 30 cm square mother board by a bonding technique.
In order to handle more than a thousand-signal lines 
in a 30 cm square size,
the structure of 8 layers and 
the micro resistor arrays were adopted.
Figure \ref{fig:f2}  shows the mother board and the gas vessel  in which 
the 10 cm square MSGC is mounted.
The preamplifier cards are inserted vertically  
to the connecter on the rear side of the  motherboard, 
which has 64 fast amplifiers (MQS104 developed by LeCroy) and discriminators.
All discriminated signals from the anodes and the cathodes (ECL level) are fed
to the position encoded system mentioned hereafter.  

As pointed out in the Tanimori et al.~[8][9],
the fast and narrow pulses from both anodes and back strips 
provide very tight timing coincidence between anodes and
back strips within $\sim$10ns.
This means that the two coordinates of an incident point  
are able to be synchronously 
encoded with a  few ten ns clock cycle by
requiring the coincidence between the timings of both anodes and back strips.
This procedure enables us to encode more than 10$^7$ events per second.
Since almost all events generates about three hit strips on both anodes 
and back strips,
a simple method of getting the hit position as a  center gravity 
of the hit electrodes can provide a position resolution of 
less than 100 $\mu$m.
This resolution reaches the limit due to the diffusion of drift electrons.
Therefore, we need to record only the positions of the hit anodes and 
back strips instead of the pulse heights of those electrodes. 

\begin{figure}
\begin{center}
\includegraphics[width=120mm]{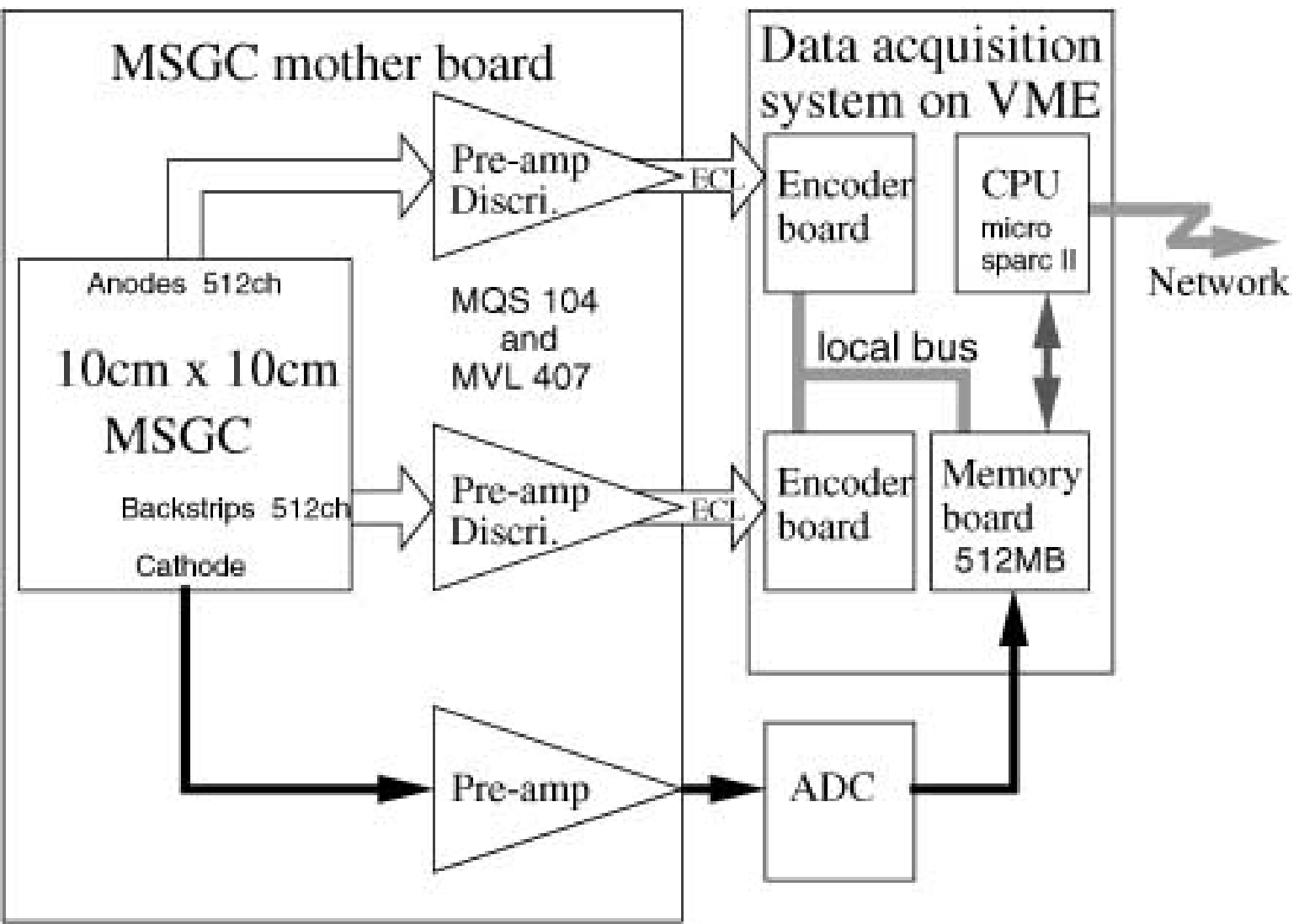}
\end{center}
\caption{}
\label{fig:f3}
\end{figure}

\begin{figure}
\begin{center}
\includegraphics[width=120mm]{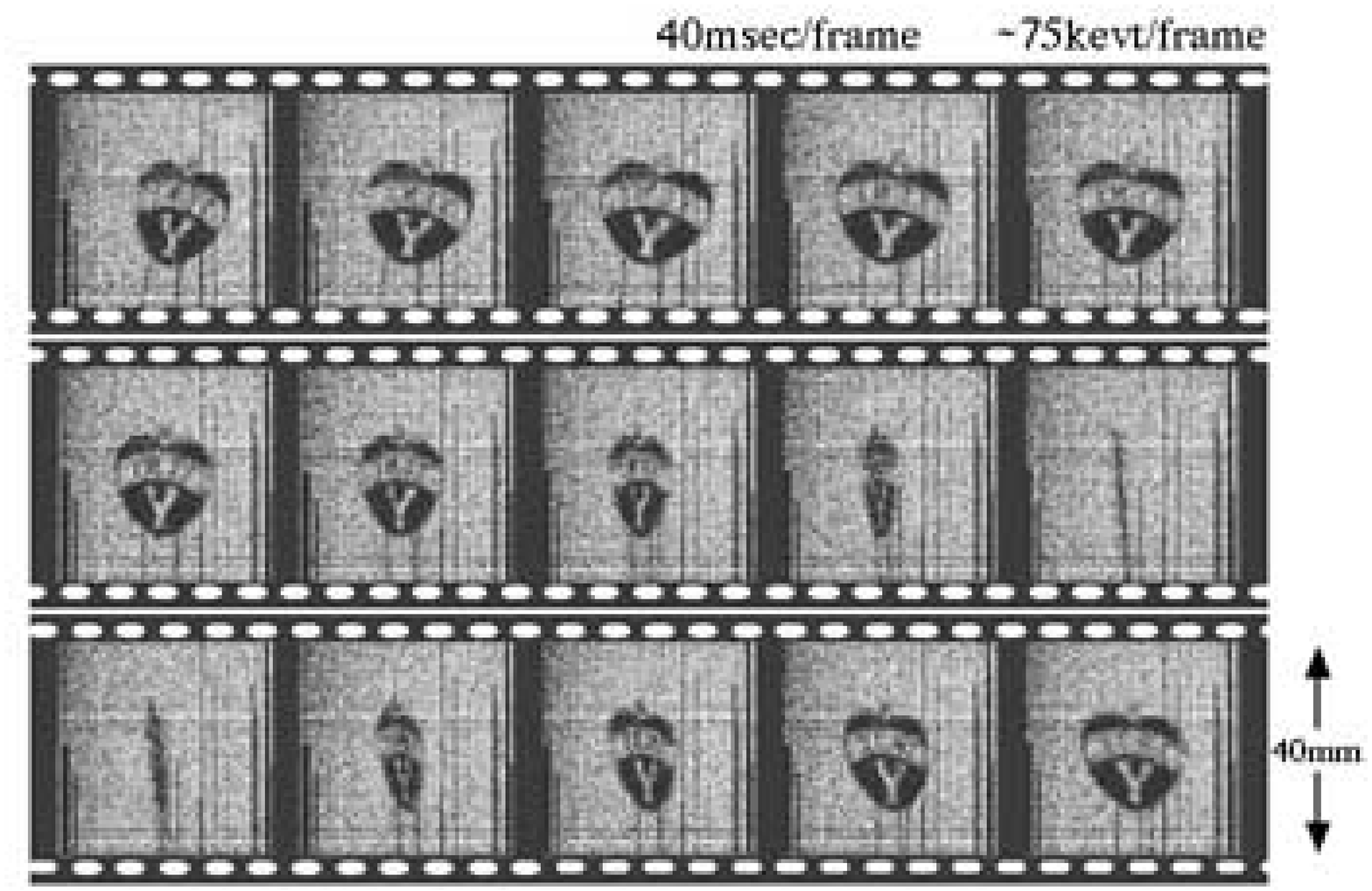}
\end{center}
\caption{}
\label{fig:f4}
\end{figure}

To realize the above idea for handling more than million events per second,
the synchronous encoding
system has been developed, of which block diagram
is shown in Fig.\ref{fig:f3}.
The readout system consists of 9-U VME modules of two types.
One is the position encoding module (PEM) which has 128 inputs and  
trees of Programmable Logic Devices (PLD).
The PEM  encodes  hit strips to X or Y coordinates. 
The other is the control and memory module (CMM) which has
a large buffer memory of $\sim$200 Mbyte for keeping the image data during 
$\sim$10 seconds at the counting rate of 10$^7$ events/s, 
and generates the synchronous clock.
The new system can handle more than 3 million  events per second, 
which is more than 3$\times$10$^3$ times the ability of the CAMAC system. 
This enables us to take $\sim$ 30 frames/seconds of 
images with enough quality. 
Figures \ref{fig:f4} 
show the several sequent frame images in the movie  
taken  a metal pendant 
rotating at the front of the MSGC by an X-ray irradiation,
where 25 images per second were taken. 

The details of the readout method and those VME modules  are 
described in Tanimori et al. [9] and Ochi et al. [3], respectively.

\section{New approach for crystal analysis}

\begin{figure}[htbp]
\begin{center}
\begin{tabular}{cc}
\includegraphics[width=80mm]{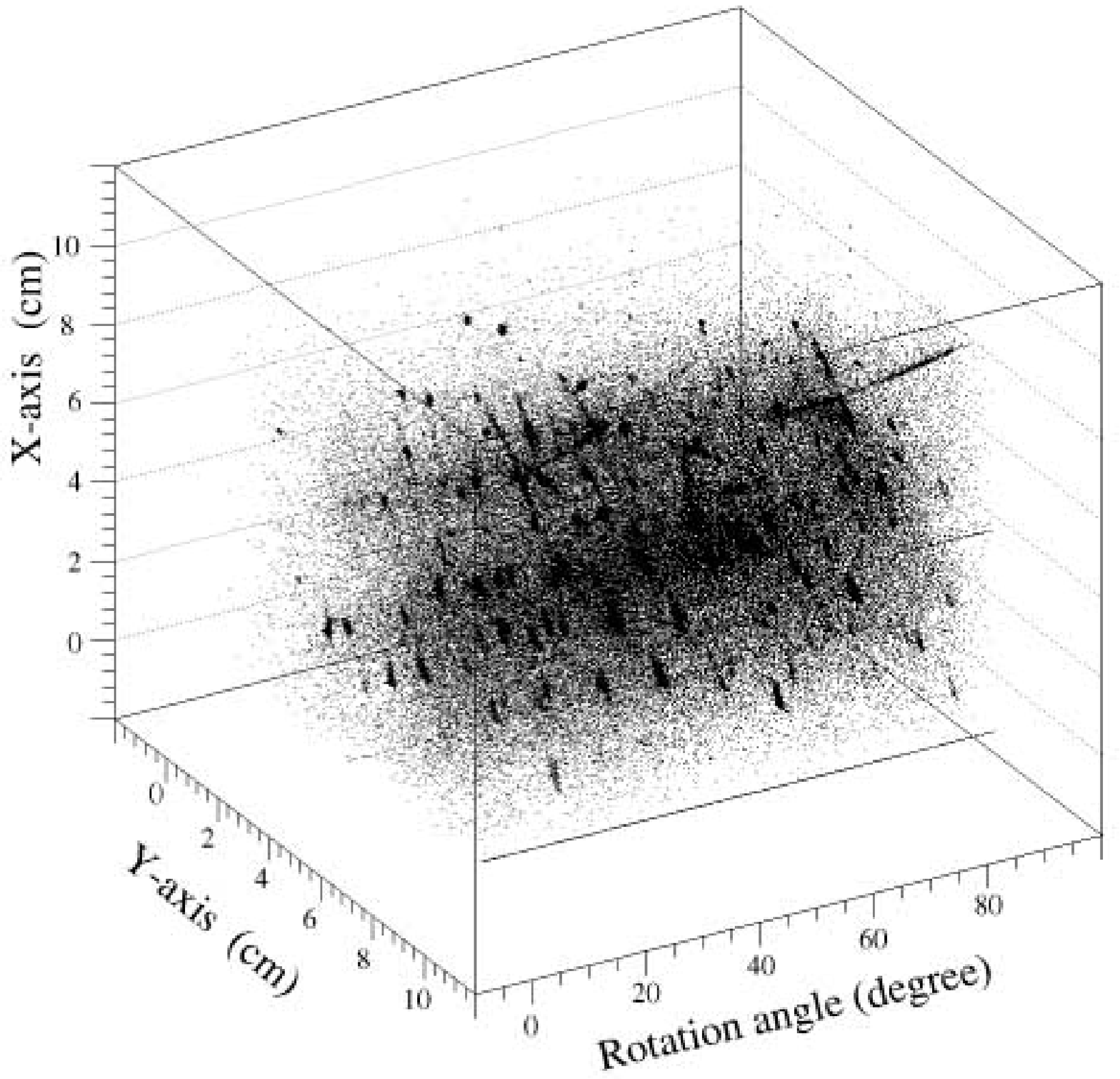} &
\includegraphics[width=80mm]{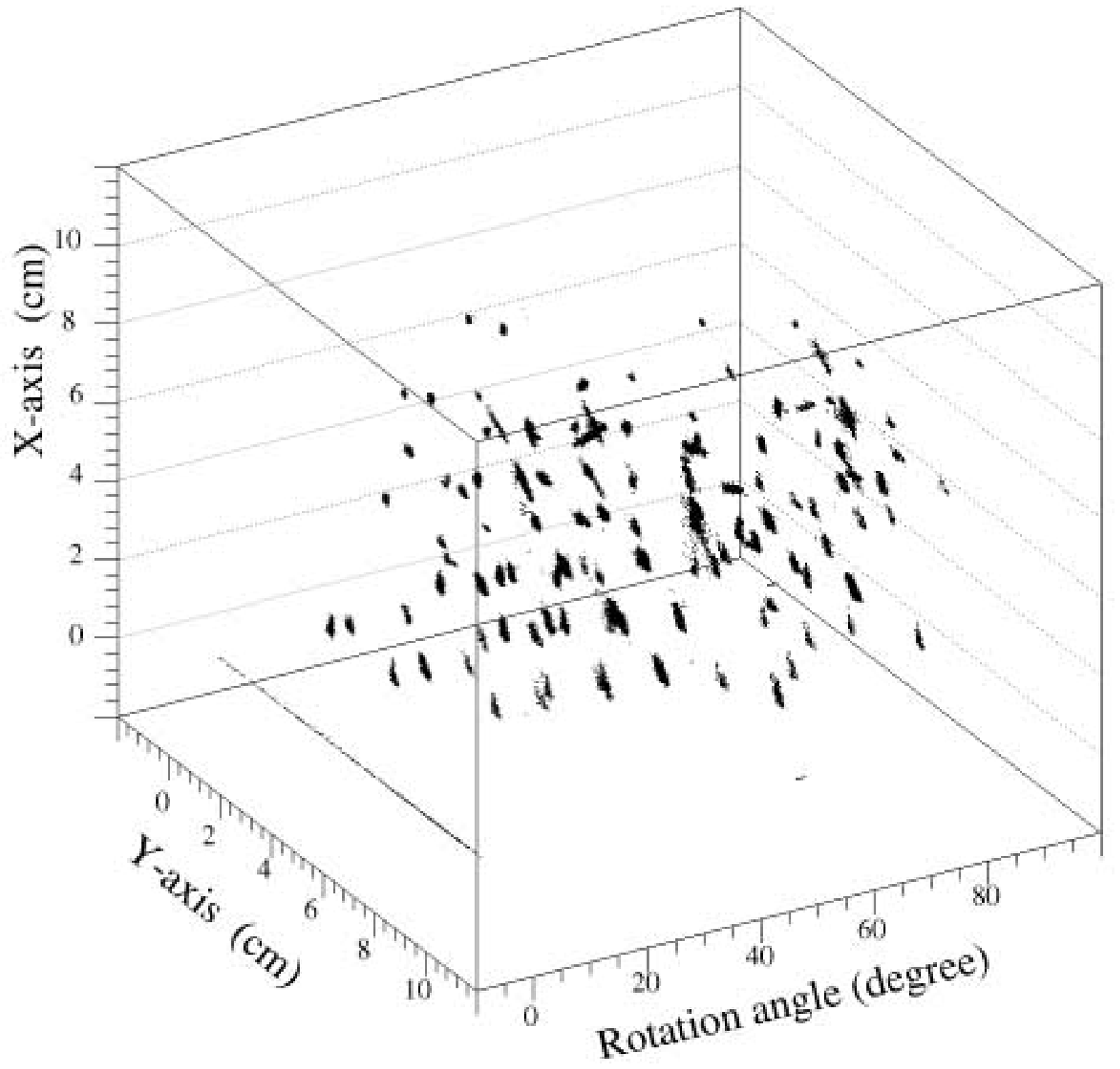} \\
(a) & (b) \\
\end{tabular}
\end{center}
\caption{}
\label{fig:f5}
\end{figure}

\begin{figure}
\begin{center}
\includegraphics[width=120mm]{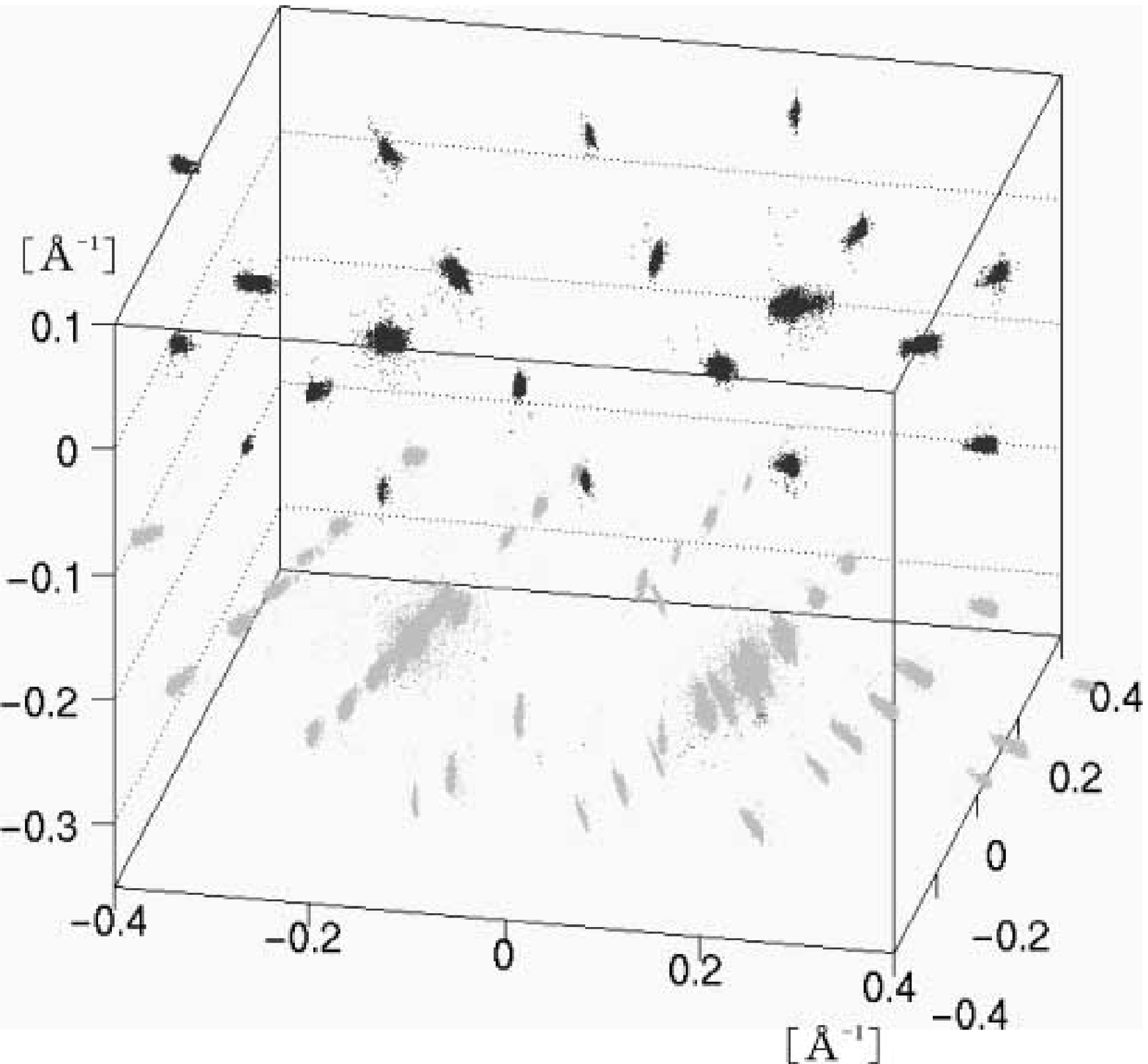}
\end{center}
\caption{}
\label{fig:f6}
\end{figure}

In general, a two-dimensional image of a diffraction pattern  
is not sufficient to obtain the three dimensional information
of the objective crystal.
When using a monochromatic X-ray beam, several diffraction patterns
are taken varying the angle between one axis of the crystal
and the X-ray beam over an acceptable  angle range (a few degrees).
The MSGC can record the arriving time 
of each X-ray photon with a few ten ns resolution.
The timing just gives us the information on the angle of the rotation 
of the crystal with a very fine angular resolution. 
Figures \ref{fig:f5}(a) shows the three-dimensional images 
obtained from MSGC
which consists of two positional and one rotating angle 
coordinates.
You note that this fine angular resolution of $\le$ 0.1 degree
is obtained for each diffraction spot,
which enable to remove the noises spreaded uniformly in this space 
from real spots made by an X-ray diffraction from a target crystal.
Using this method, the sensitivity for a faint diffraction spot can be 
improved more than 10 times as shown in Fig.\ref{fig:f5}(b). 
For crystals having the little constituent atoms,
the MSGC allows us to get all the information needed for 
crystal analysis from only a few minutes measurement with one
continuous rotation of a crystal.
Fig.\ref{fig:f6} shows a reciprocal lattice image calculated from the 
data obtained by the MSGC.
Thus the MSGC would dramatically improve an X-ray crystallography.

The most intriguing approach for an X-ray crystallography
using an ultimate time resolved image is a direct observation 
of the dynamical change of a crystal structure
for periodic variations or reactions.
The imaging device based on the photo counting method such as MWPC or MSGC
has an essential upper limit of $\le$ 10$^7$ events/s for handling
the data, which restricts the number of picture frames to less than 
hundreds per second.
However, a fast process within $\le$ $\mu$s 
can be observed as continuous images    
if periodic measurements are done for this process.
Since the MSGC records the timings both of each detected X-ray
and of each periodic process, 
all X-rays obtained by the MSGC can be folded into one phase of the
periodic process.
When a process with variation times of 100 $\mu$s
is measured periodically by the MSGC 
at the event rate of a few MHz during 10s,
hundred images with $\sim$1 $\mu$s timing bin  
could be obtained in one process.
Each image made of about 10$^6$ X-rays gives
a high quality picture.
We already applied this method,
and succeeded to catch the dynamical change of the crystal structure
of [Bu$_4$N]$_4$[Pt$_2$(pop)$_4$] between a photon excited state
and a stable state first in the world.

Details of the X-ray crystallography application 
and another  potentialities of MSGC are described in Ochi [11].

\section{Diagnosis of discharges: Capillary intermediate multiplier }

Although the technology of a MSGC seems to be established due to
the recent intensive studies in the world,
there still remains one crucial problem to prevent a MSGC
from the stable operation: discharges damage the electrodes of a MSGC.
The process of the discharge in MSGCs are studied 
in detail by Peskov, Ramsey \& Fonte [5].
Although we do not know the complete diagnosis of discharges in a MSGC yet,
the tolerance can be increased by several improvements.
About 5$\sim$8\% of electrodes were damaged by discharges during one 
years operation in our experience, testing lots of 5 cm square MSGC.
Since damage due to discharges usually concentrated in warming up 
at the first use, 
dust and  parts of bad quality of electrodes might be the reasons for
discharges.
For the 10cm square MSGC, the surface  are now looked into by a microscope 
before the operation to remove dusts.
Also chromium has been used as the material of the electrode in the latest MSGCs
due to its higher melting point.
These efforts have distinctly suppressed the occurrences of 
broken strips by a discharge.

Another solution is the insertion of 
an intermediate gas-multiplier such as Gas Electron Multiplier (GEM)
proposed by Bouclier et al.~[1].
An intermediate gas-multiplier was at first realized in a multi-step avalanche
chamber using fine mesh planes,
which were intensively studied around 1980. 
Originally an intermediate gas-multiplier was used to attain 
a very high gain by combined with a MWPC, 
for detecting one ultraviolet photon of Cherenkov light.
We are now investigating a capillary plate as an 
intermediate gas-multiplier, which consists of 
a bundle of fine glass capillaries 
with uniform length, the ends of which forms flat planes and
was coated by Inconel metal.
The gas multiplication of  a capillary plate in gases 
has already been confirmed 
by Sakurai et al.~[6]. 
In this paper, high voltages 
were fed to both end planes of a capillary plate with
the diameter of 2cm,
and  the high-electric field in a capillary induced a gas multiplication.
Its gain was reported to reach up to more thousands.
The detail is mentioned in this reference.

\begin{figure}[htbp]
\begin{center}
\includegraphics[width=160mm]{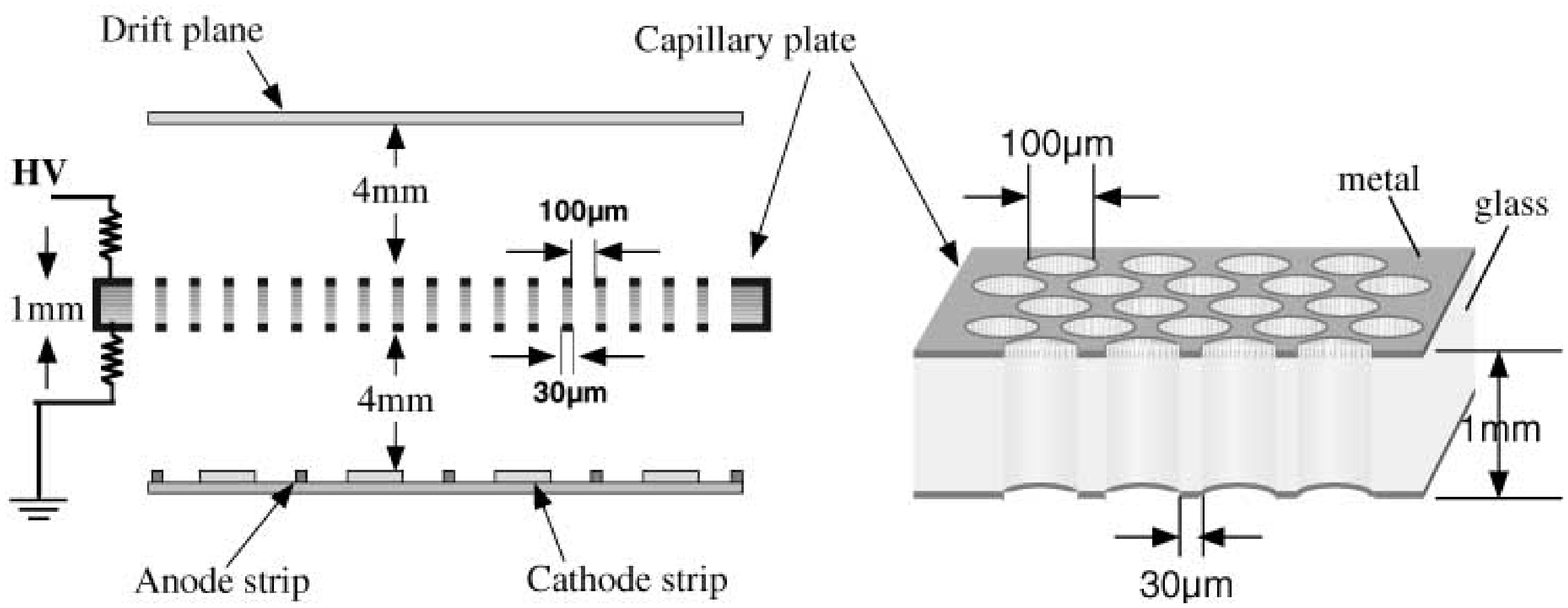}
\end{center}
\caption{}
\label{fig:f7}
\end{figure}

Figure \ref{fig:f7} shows the side view of our system 
combined with the 10cm square capillary 
and  the 10 cm square MSGC, where capillary are set 4mm above the MSGC. 
The large capillary plate used here is made by Hamamatsu Photonics,
and the diameter and the length of its capillary are 100 $\mu$m and 1mm
respectively.  
A thickness of a capillary plate of $\sim$ 1mm, which is  more ten times
thicker than that of GEM, and the very high surface resistivity 
of a capillary easily let us infer the unstable operation of its
gas multiplication under a high intense radiation
due to a space-charge effect in a capillary.
Actually, non-uniformity and instability of a gain were observed  every
measurement for this 10 cm square capillary even under a relative
low irradiation of $\sim $ 100 Hz/mm$^2$.
Under a medium radiation, most bright parts in a image obtained by this 
system were observed to diminish soon. 
By such instability, we could not evaluate the performance of it 
quantitatively. 

\begin{figure}[htbp]
\begin{center}
\includegraphics[width=160mm]{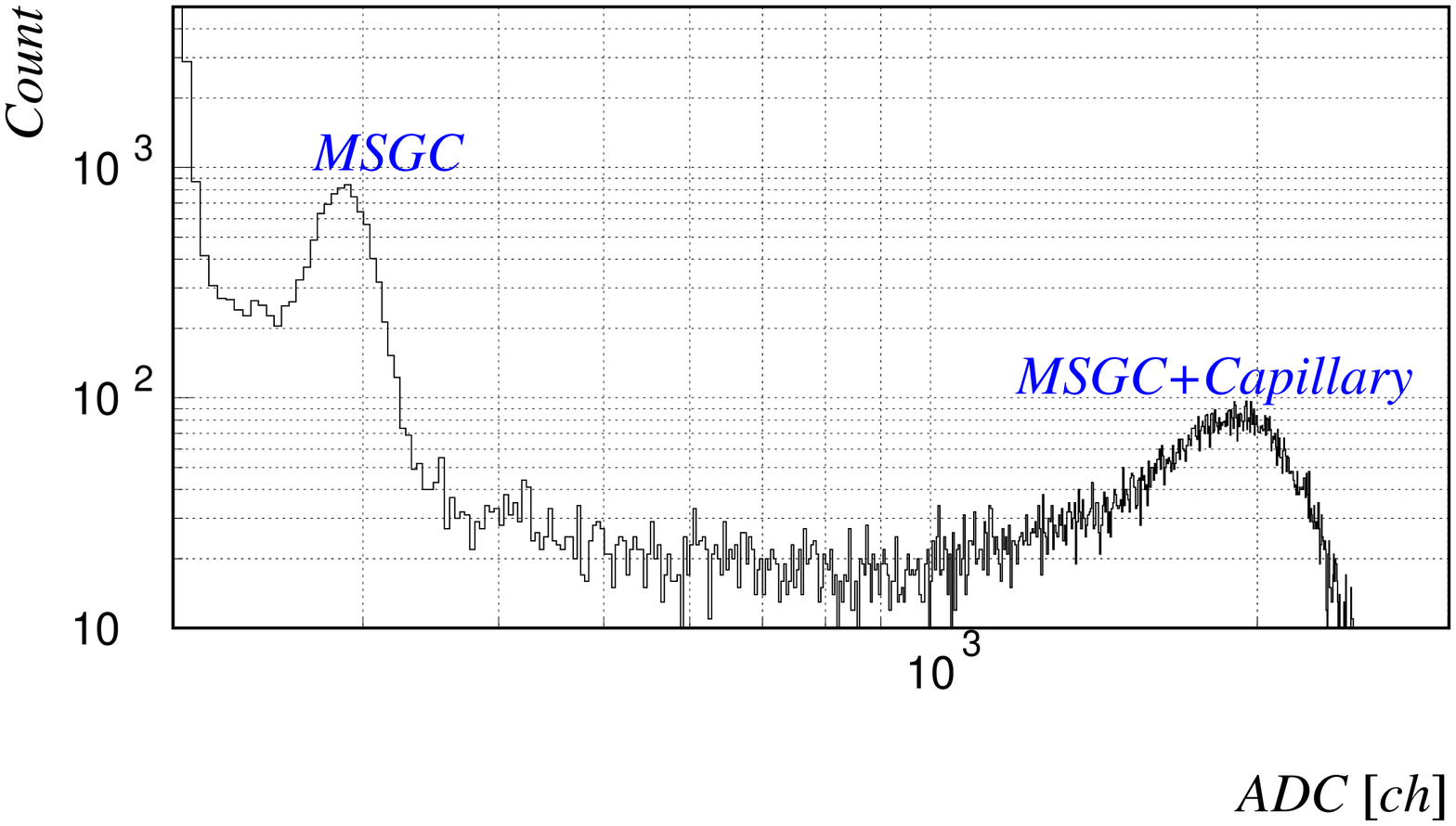}
\end{center}
\caption{}
\label{fig:f8}
\end{figure}

\begin{figure}[htbp]
\begin{center}
\includegraphics[width=160mm]{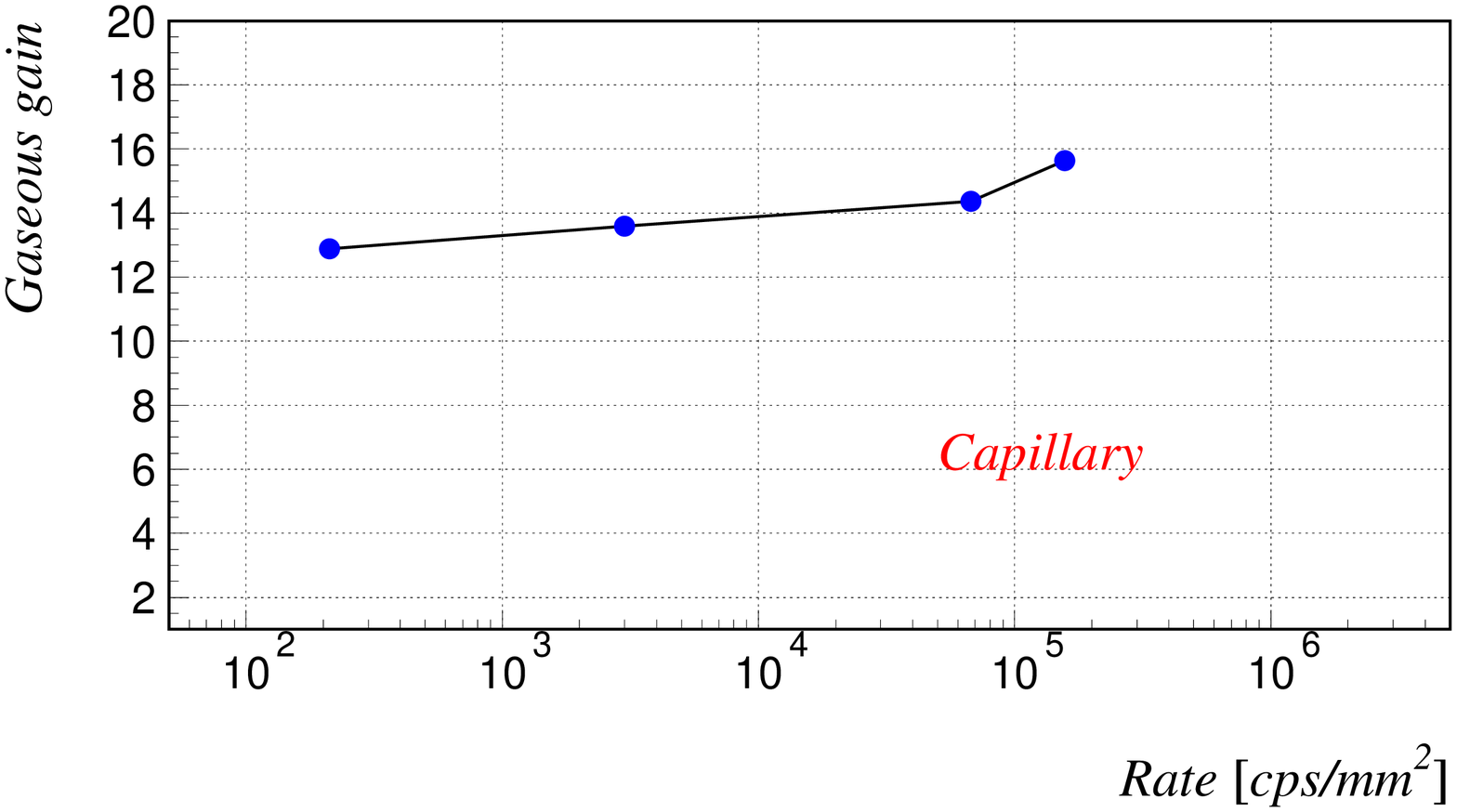}
\end{center}
\caption{}
\label{fig:f9}
\end{figure}

\begin{figure}[htbp]
\begin{center}
\begin{tabular}{cc}
\includegraphics[width=80mm]{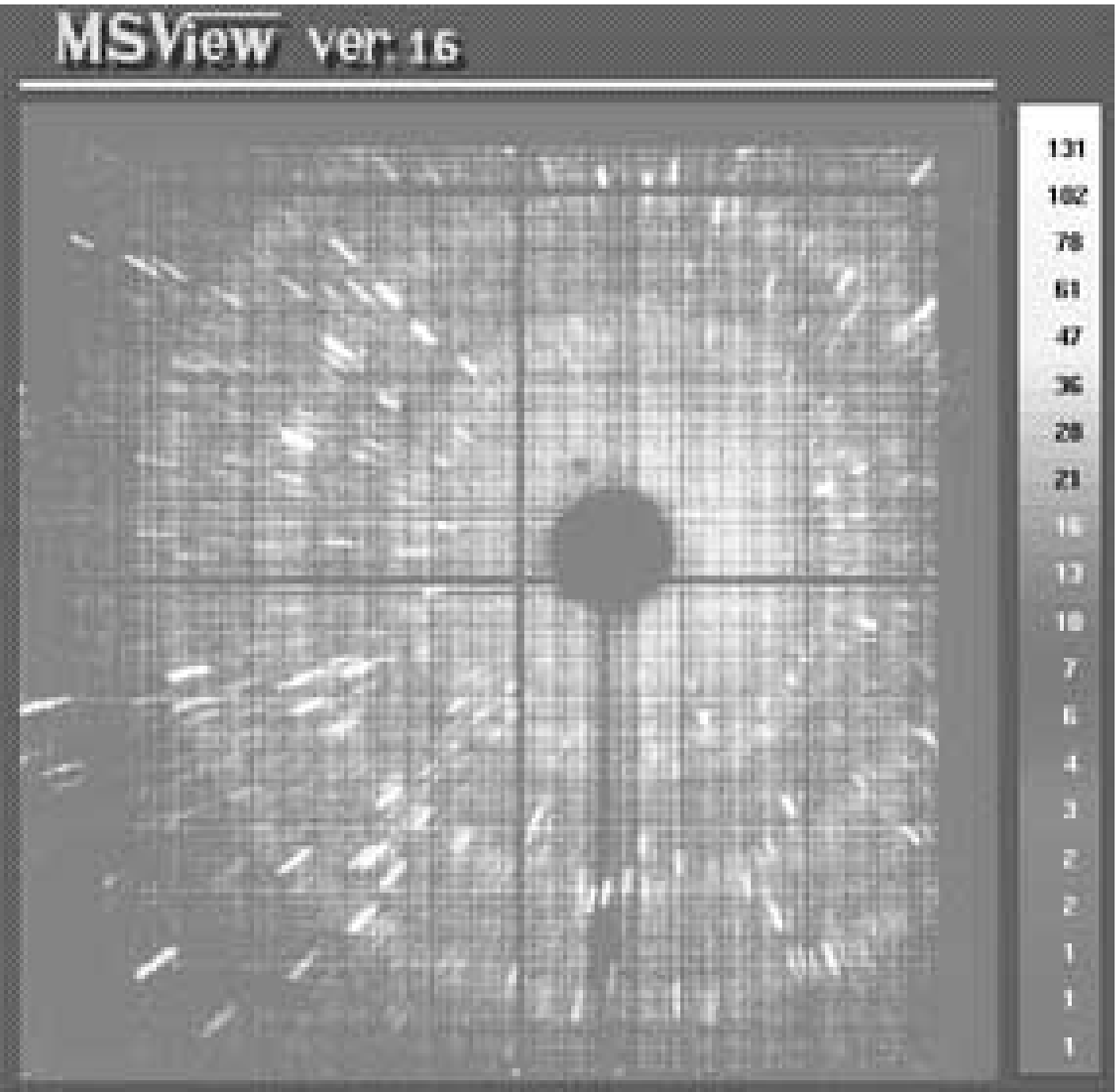} &
\includegraphics[width=80mm]{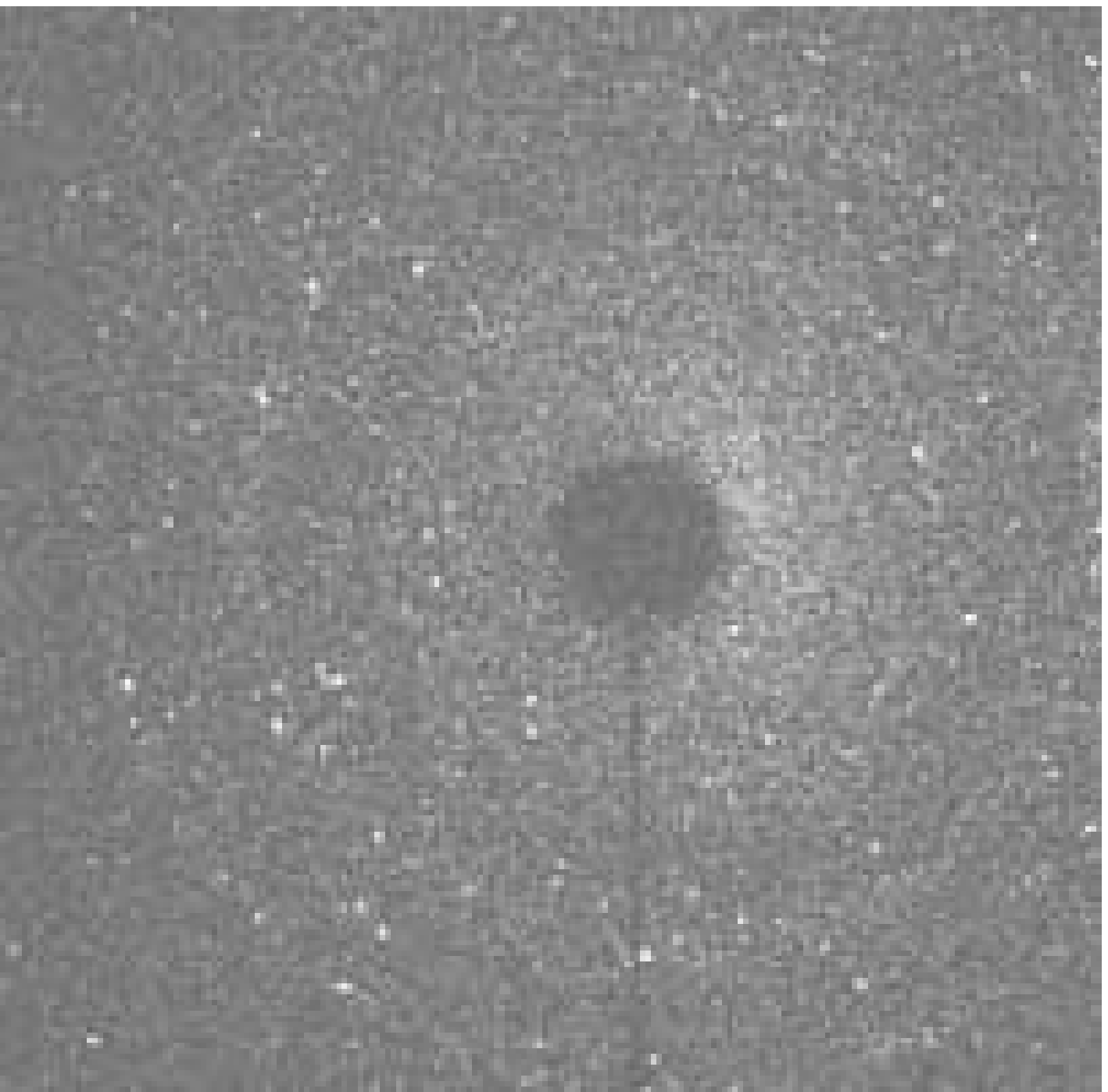} \\
(a) & (b) \\
\end{tabular}
\end{center}
\caption{}
\label{fig:f10}
\end{figure}

In order to absorb ions in a capillary, 
a little conductivity was added to  the surface of the capillary,
by which the resistivity of 40 M$\Omega$ appeared 
between both sides of the 10cm square capillary.
This conductive capillary has dramatically improved the
performance of this system.
Figure \ref{fig:f8} shows the energy spectrum of Cu characteristic X-rays 
obtained by this improved system, 
where the peak generated by a single MSGC and that by combined MSGC and capillary
plates are obviously distinguished.
From this figure the gain of the capillary itself can be estimated,
and the rate capability was measured. 
As shown in Fig.\ref{fig:f9}, 
the capillary was observed to be operated steadily 
up to  more than a 10$^5$Hz/mm$^2$.
The gain of a conductive capillary reached more than $\sim$ 3000, 
and non-uniformity of the gain more than 10\% was not  observed. 
Figures \ref{fig:f10}(a) and (b) show the comparison of the image performance 
between this system and Imaging Plate using the powder diffraction of sugar 
for same exposure times;  the very good performance of this system
are distinctly noted. 
Here this system was operated with the total gain of 1000
(Cu characteristic X-rays were used),
and the gain of the MSGC itself was only a few tens. 
By adopting this intermediate multiplier, 
the total gain was increased about ten times, and
the operation voltage of the MSGC between anodes and cathodes
could be reduced by $\sim$ 100 V.
This condition ensures the stable operation free from both discharges
and electrical noises even under an intense irradiation. 
The detail of the study on the conductive capillary plate will be described in 
Nishi et al. [10]. 

\noindent{\large \bf Summary}

We have developed a two-dimensional MSGC and 
the fast readout system,
both of which are essential developments to realize a
new time resolved X-ray imaging detector.
In addition,
the new type of an intermediate electron multiplier
has been proposed: 
the capillary plate of which capillary has a conductive
surface was made and set above the MSGC.
This combined system was found to be operated very stably
with an enough gain, which has never
been attained by a simple MSGC.
Furthermore, supplied voltages for anodes and cathodes
of the MSGC can be kept  within the quite safety range against
discharges, and 
we have been free from the risk of the
destruction of electrodes due to  discharges.  
Such an desirable operation of the MSGC 
provides an ideal image having very high qualities
such as good position resolution, no distortion, 
very wide-dynamic range, 
and steady and flat uniformity of the efficiency;
the new application  of this device for the 
X-ray imaging analyses has been able to 
be discussed with reality as described in above section.

We also stress that this detector system is a complete  
electric system controlled by the computer, 
where electric system means not only that the data are
electrically transferred to computers,
but also that all the components of this system are made
from IC  technology.
The MSGC itself is made using high density printed board
technology for the direct mounting of a bare LSI.
All electrical elements of the MSGC readout system 
are made of commercially available LSI chips.    
In this system,
IC chips with single function such as 
fast amplifiers, comparators, and
the ECL-TTL transfers  are the main elements of
the components,
whereas PLDs, the core parts of the readout system, occupies less than
10\% of the system.
Then  we  have begun to  redesign the readout system using commercially
manufactured  IC chips having 32channels amplifiers or discriminators 
in one chip which
will be mounted directly on the MSGC mother board.
PLDs for the position encoding also will be set in the  
the MSGC box, and no ECL-TTL transfer chip is needed. 
This improvement will realize
a handy  MSGC imager similar to a liquid crystal display 
in very near future.


\begin{ack}

We gratefully acknowledge the kind support and 
fruitful discussion of Prof. Y.~Ohashi, Dr. H.~Uekusa 
and their colleagues
of Dept. of Chemistry, Tokyo Institute of Technology.
T.~Tanimori, Yuji Nishi, and A.~Ochi would like to thank 
Dr. T.~Ueki, Dr. M.~Suzuki, Dr. T.~Fujisawa, Dr.~Toyokawa and
members of Biological Physics group of 
The Institute of Physical and Chemical Research (RIKEN)
and Japan Synchrotron Radiation Research Institute (JASRI)
for the continuous support and encouragement. 
This work is supported by 
CREST:Japan Science and Technology Corporation (JST)
and partially by JASRI. 

\end{ack}

\newpage
\noindent{\large \bf Figure Captions}

\vspace{3mm}
\noindent{\bf Figure 1.} 
Schematic structure of the two-dimensional 
MSGC which  were formed  on a 17 $\mu$m  thin 
polyimide substrate.
On the polyimide layer, 7 $\mu$m wide anodes  and 63 $\mu$m wide cathodes
were formed
with a 200 $\mu$m pitch
(the width of the cathodes was changed to 100 $\mu$m 
in the imaging measurement as mentioned in section 4). 
Between the ceramic base and the polyimide substrate,
there are set back strips with a 200 $\mu$m pitch
orthogonal to the anodes. 
All electrodes are made of gold with a thickness of 1 $\mu$m
(recently chromium are used as mentioned in section 5).
To define the drift field, the drift  plane was placed at 10 mm above 
the substrate.
                        
\vspace{2mm}
\noindent{\bf Figure 2.} 
Top view of the new 10 cm square MSGC mother board.
The MSGC is mounted in the gas vessel seen in 
the lower right of the mother board. 

\vspace{2mm}
\noindent{\bf Figure 3.} 
Block diagram of the new synchronous readout system.

\vspace{2mm}
\noindent{\bf Figure 4.} 
Several sequent frame images in the movie  
taken  a metal pendant 
rotating at the front of the MSGC bye an X-ray irradiation,
where 25 images per second were taken.

\vspace{2mm}
\noindent{\bf Figure 5.} 
(a)Three dimensional image (X, Y and rotating angle coordinates) 
of X-ray diffraction spots
of the Phenothiazine-Benzilic acid complex,
in which the sample crystal at 10 cm front of the MSGC 
was rotated continuously along the axis normal to 
incident monochromatic X-ray beam.
(b)similar image after the noise reduction mentioned in the text.

\vspace{2mm}
\noindent{\bf Figure 6.}
Reciprocal lattice image obtained by the MSGC.

\vspace{2mm}
\noindent{\bf Figure 7.} 
Side view of the 10cm square capillary 
and  the 10 cm square MSGC.

\vspace{2mm}
\noindent{\bf Figure 8.} 
Energy spectrum of Cu characteristic X-rays 
obtained by the 10cm square capillary 
and  the 10 cm square MSGC,
where the peak generated by a single MSGC and that by combined MSGC and capillary
plates are obviously distinguished

\vspace{2mm}
\noindent{\bf Figure 9.} 
Rate Capability of the conductive capillary plate.

\vspace{2mm}
\noindent{\bf Figure 10.}
Comparison of the image performance 
between the MSGC + conductive capillary (a)
and Imaging Plate (b) using the powder diffraction of sugar 
for same exposure times.  

\end{document}